# OVERVIEW OF THE NEW HORIZONS SCIENCE PAYLOAD

H. A. Weaver[a], W. C. Gibson[b], M. B. Tapley[b], L. A. Young[c], and S. A. Stern[c]
[a]Johns Hopkins University Applied Physics Laboratory, 11100 Johns Hopkins Road, Laurel, MD 20723
[b]Southwest Research Institute, 6220 Culebra Road, San Antonio, TX 78238
[c]Southwest Research Institute, 1050 Walnut St., Suite 400, Boulder, CO 80302

## Abstract

The New Horizons mission was launched on 2006 January 19, and the spacecraft is heading for a flyby encounter with the Pluto system in the summer of 2015. The challenges associated with sending a spacecraft to Pluto in less than 10 years and performing an ambitious suite of scientific investigations at such large heliocentric distances (> 32 AU) are formidable and required the development of lightweight, low power, and highly sensitive instruments. This paper provides an overview of the New Horizons science payload, which is comprised of seven instruments. *Alice* provides moderate resolution (~3-10 Å FWHM), spatially resolved ultraviolet (~465-1880 Å) spectroscopy, and includes the ability to perform stellar and solar occultation measurements. The *Ralph* instrument has two components: the *Multicolor Visible Imaging Camera (MVIC)*, which performs panchromatic (400-975 nm) and color imaging in four spectral bands (Blue, Red, $CH_4$, and NIR) at a moderate spatial resolution of 20 μrad/pixel, and the *Linear Etalon Imaging Spectral Array (LEISA)*, which provides spatially resolved (62 μrad/pixel), near-infrared (1.25-2.5 μm), moderate resolution ($\lambda/\delta\lambda$ ~ 240-550) spectroscopic mapping capabilities. The *Radio Experiment (REX)* is a component of the New Horizons telecommunications system that provides both radio (X-band) solar occultation and radiometry capabilities. The *Long Range Reconnaissance Imager (LORRI)* provides high sensitivity (V < 18), high spatial resolution (5 μrad/pixel) panchromatic optical (350-850 nm) imaging capabilities that serve both scientific and optical navigation requirements. The *Solar Wind at Pluto (SWAP)* instrument measures the density and speed of solar wind particles with a resolution $\Delta E/E < 0.4$ for energies between 25 eV and 7.5 keV. The *Pluto Energetic Particle Spectrometer Science Investigation (PEPSSI)* measures energetic particles (protons and CNO ions) in 12 energy channels spanning 1-1000 keV. Finally, an instrument designed and built by students, the *Venetia Burney Student Dust Counter (VB-SDC)*, uses polarized polyvinylidene fluoride panels to record dust particle impacts during the cruise phases of the mission.

## 1. Introduction

New Horizons was the first mission selected in NASA's New Frontiers series of mid-sized planetary exploration programs. The New Horizons spacecraft was launched on 2006 January 19 and is now on a 3 billion mile journey to provide the first detailed reconnaissance of the Pluto system during the summer of 2015. Assuming that this primary objective is successful, NASA may authorize an extended mission phase that will permit a flyby of another Kuiper belt object (KBO), as yet unidentified, probably within 3 years of the Pluto encounter. The genesis and development of the New Horizons mission is described by Stern (2007). The scientific objectives of the mission are discussed by Young et al. (2007). Here we provide a high level overview of the scientific payload. Detailed descriptions of individual instruments are given elsewhere in this volume, as referenced below.

The New Horizons mission is an ambitious undertaking that required the development of lightweight, low power, and highly sensitive instruments. Pluto will be nearly 33 AU from the sun at the time of the encounter in 2015, and a launch energy (C3) of nearly 170 $km^2$ $s^{-2}$ was needed to reach this distance within the 9.5 year transit to the Pluto system. Even using the powerful Lockheed-Martin Atlas 551 launcher in tandem with its Centaur second stage and a Boeing Star48 third stage, the entire spacecraft mass had to be kept below 480 kg, of which less than 50 kg was allocated to the science payload. At Pluto's large heliocentric distance, the use of solar photovoltaic cells was not an option, so the New Horizons mission relies on a radioisotope thermoelectric generator (RTG) for all of its power needs. The mission requirement on the total power available at the Pluto encounter is only 180 W, of which less than 12 W can be used at any one time by the scientific instruments. The solar output (light and particle) at Pluto is approximately 1000 times smaller than at the Earth, which means that the instruments attempting to measure reflected sunlight or the solar wind during the Pluto encounter must be extremely sensitive. Finally, we note that the long mission duration imposes strict reliability requirements, as the spacecraft and science payload must meet their performance specifications at least 10 years after launch.

Fortunately, all of the New Horizons instruments successfully met these daunting technical challenges without compromising any of the mission's original scientific objectives. Below we provide a high-level description of all the instruments on New Horizons, discuss their primary measurement objectives, and summarize their observed performance, which has now been verified during in-flight testing. But first we begin by briefly describing the spacecraft pointing control system as it relates to the science payload.

## 2. Payload Pointing Control

The New Horizons spacecraft does not have enough power to support a reaction wheel based pointing control system and instead relies on hydrazine thrusters to provide slewing capability and attitude control. The positions of stars measured by one of two star trackers (the second star tracker provides redundancy) are used to determine the absolute orientation of the spacecraft (i.e., the RA and DEC locations of some reference axis on the spacecraft), and the drift rate is monitored by a laser-ring gyro system (the inertial measurement unit, or IMU). The attitude data from the star tracker and IMU are used in a feedback loop to set the pointing within prescribed limits in both absolute position and drift rate. The spacecraft IMUs, star trackers, sun sensors, and guidance computers are all redundant.

The New Horizons spacecraft spends much of its time spinning at ~5 RPM around the Y-axis. In this mode, useful data can be obtained by REX, SWAP, PEPSSI, and the VB-SDC, but typically not by any of the other instruments.

For virtually all observations made by the imaging instruments, 3-axis pointing control mode is required. In 3-axis mode, the spacecraft can be slewed to a targeted location to an accuracy of ±1024 µrad (3σ) and controlled to that location within a typical "deadband" of ±500 µrad. For some Alice observations, when the target must be kept near the center of its narrow slit, the deadband can be reduced to ±250 µrad. The drift rate is controlled to within ±34 µrad/sec (3σ) for both fixed and scanning observations. The post-processing knowledge of the attitude and drift rate derived from the star tracker and IMU data are ±350 µrad (3σ) and ±7.5 µrad/sec (3σ), respectively. Ralph observations usually require the spacecraft to scan about its Z-axis. The nominal scan rate for Ralph/MVIC is 1.1mrad/sec, and the nominal scan rate for Ralph/LEISA is 0.12 mrad/sec.

Further details about the New Horizons guidance and control system can be found in Rogers et al. (2006).

# 3. Science Payload

## 3.1 OVERVIEW

All of the fundamental ("Group 1") scientific objectives for the New Horizons mission (Stern 2007; Young et al. 2007) can be achieved with the *core* payload comprised of: (i) the *Alice* ultraviolet (UV) imaging spectroscopy remote sensing package, (ii) the *Ralph* visible and infrared imaging and spectroscopy remote sensing package, and (iii) the *Radio Experiment (REX)* radio science package. The *supplemental* payload, which both deepens and broadens the mission science, is comprised of the *Long Range Reconnaissance Imager (LORRI)*, which is a long-focal-length optical imaging instrument, and two plasma-sensing instruments: the *Solar Wind Around Pluto (SWAP)* and the *Pluto Energetic Particle Spectrometer Science Investigation (PEPSSI)*. The supplemental payload is not required to achieve minimum mission success, but these instruments provide functional redundancy across scientific objectives and enhance the scientific return by providing additional capabilities not present in the core payload. The *Venetia Burney Student Dust Counter (VB-SDC)*, which was a late addition to the supplemental payload approved by NASA as an Education and Public Outreach (EPO) initiative, also provided a new capability to New Horizons, namely, an interplanetary dust detection and mass characterization experiment.

Drawings of all seven instruments are displayed in Figure 1, which also lists the mass and power consumption of each instrument. The locations of the instruments on the New Horizons spacecraft are displayed in Figure 2.

As discussed further below, Ralph is essentially two instruments rolled into a single package: the *Multispectral Visible Imaging Camera (MVIC)* is an optical panchromatic and color imager; the *Linear Etalon Imaging Spectral Array (LEISA)* is an infrared imaging spectrometer. The boresights of MVIC, LEISA, LORRI, and the Alice airglow channel are aligned with the spacecraft –X axis (Fig. 2) except for minor tolerancing errors. The projections of the fields of view of those instruments onto the sky plane are depicted in Figure 3.

The types of observations performed by the New Horizons instruments are depicted in Figure 4. None of the instruments have their own scanning platforms, so the entire spacecraft must be maneuvered to achieve the desired pointings. As described below, the guidance and control system uses hydrazine thrusters to point the spacecraft at the desired target.

The principal measurement objectives and the key characteristics of the New Horizons science payload are summarized in Table I, which also includes the names and affiliations of the instrument Principal Investigators (PIs) and the primary builder organization for each instrument. The measurement objectives that are directly related to the mission Group 1 scientific objectives are highlighted in boldface. In the following subsections, we provide further discussion of each of the New Horizons instruments.

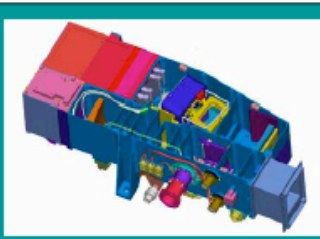
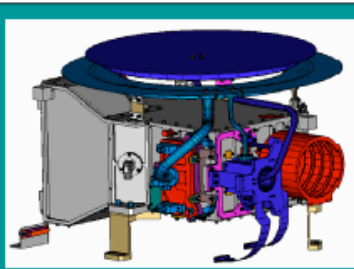
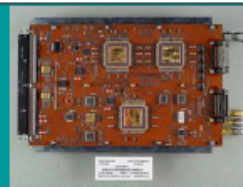

Alice UV Spectrograph, 520-1870 A, CBE mass 4.15 kg, power 4.0 W

Ralph visible & IR imager, CBE mass 10.67 kg, power 5.3 W

Radio Science Experiment (REX). CBE mass 0.1 kg, power 2.1 W

**Core Payload**

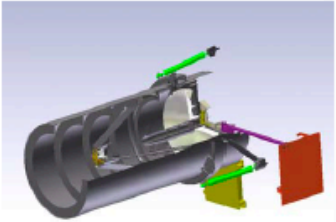
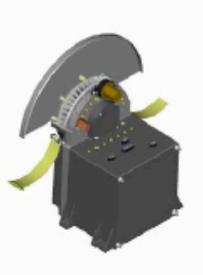
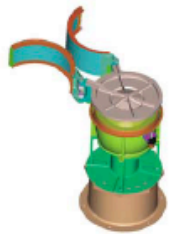
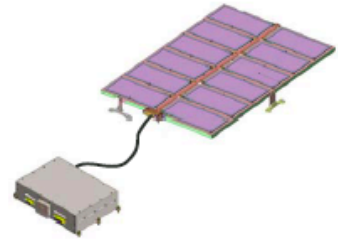

Long Range Reconnaissance Imager (LORRI), panchromatic imager. CBE mass 8.59 kg, power 5.1 W

Pluto Energetic Particle Spectrometer Science Investigation (PEPSSI). CBE mass 1.41 kg, power 2.32 W

Solar Wind Around Pluto (SWAP), CBE mass 2.94 kg, power 2.25 W

Student Dust Counter (SDC), CBE mass 1.76 kg, power 6.5 W

Fig 1: The three instruments comprising the New Horizons core payload are shown along the top row, and the instruments comprising the supplemental payload are displayed along the bottom row. The approximate mass and power consumption are shown just below the picture of each instrument. The total mass of the entire science payload is less than 30 kg, and the total power drawn by all the instruments is less than 30 W.

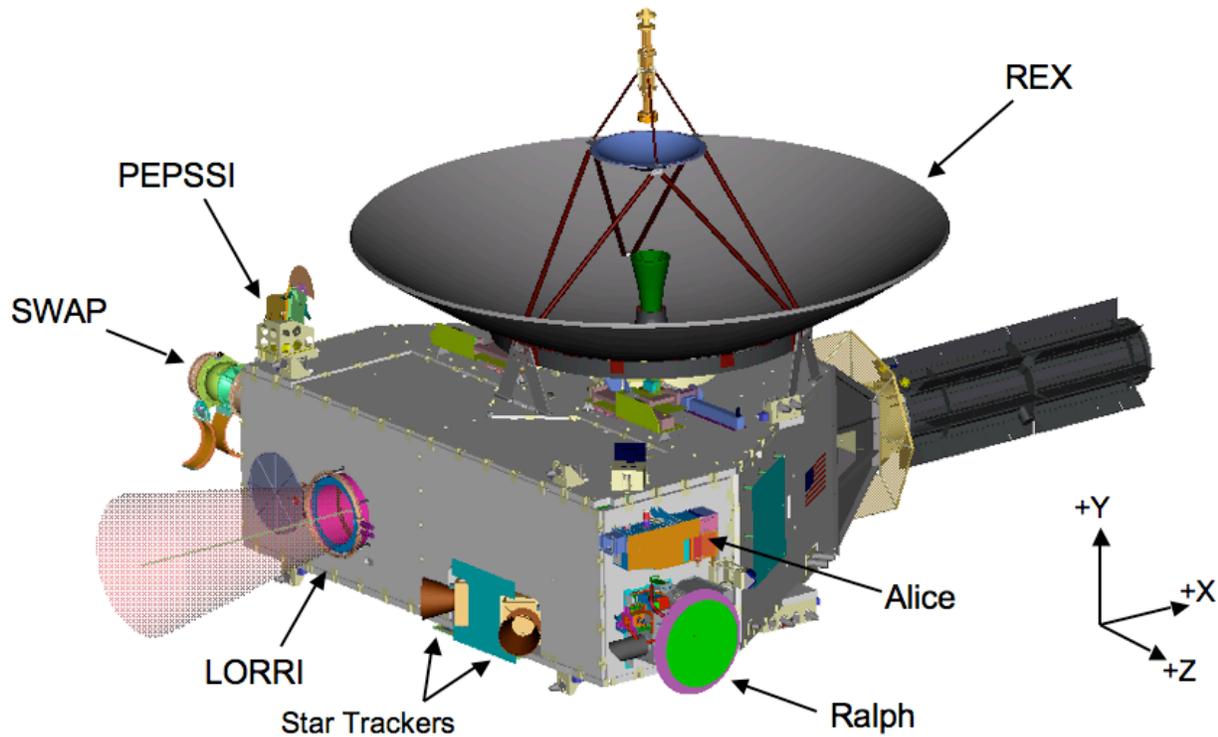

Fig. 2: This drawing shows the locations of the instruments on the New Horizons spacecraft. The VB-SDC is mounted on the bottom panel, which is hidden from view. The boresights of LORRI (sketched in figure), Ralph, and the Alice airglow channel are all approximately along the –X direction. The boresights of the Alice solar occultation channel and the antenna are approximately along the +Y direction. SWAP covers a swath that is ~200° in the XY plane and ~10° in the YZ plane. PEPSSI's field-of-view is a ~160° by ~12° swath whose central axis is canted with respect to the principal spacecraft axes to avoid obstruction by the backside of the antenna. The black structure with fins located at +X is the RTG, which supplies power to the observatory. The star trackers, which are used to determine the attitude, can also be seen. The antenna diameter is 2.1 m, which provides a scale for the figure.

Fig. 3: The fields of view (FOVs) of the MVIC, Ralph, Alice airglow, and LORRI instruments are projected onto the sky plane; the listed boresights are measured in-flight values. The angular extent of each instrument's FOV is also listed. The spacecraft +X direction is out of the page, the +Y direction is up, and the +Z direction is to the left. The LORRI field FOV overlaps the narrow portion of the Alice airglow channel, and the MVIC FOV overlaps the wide portion. The LEISA FOV overlaps the MVIC FOV.

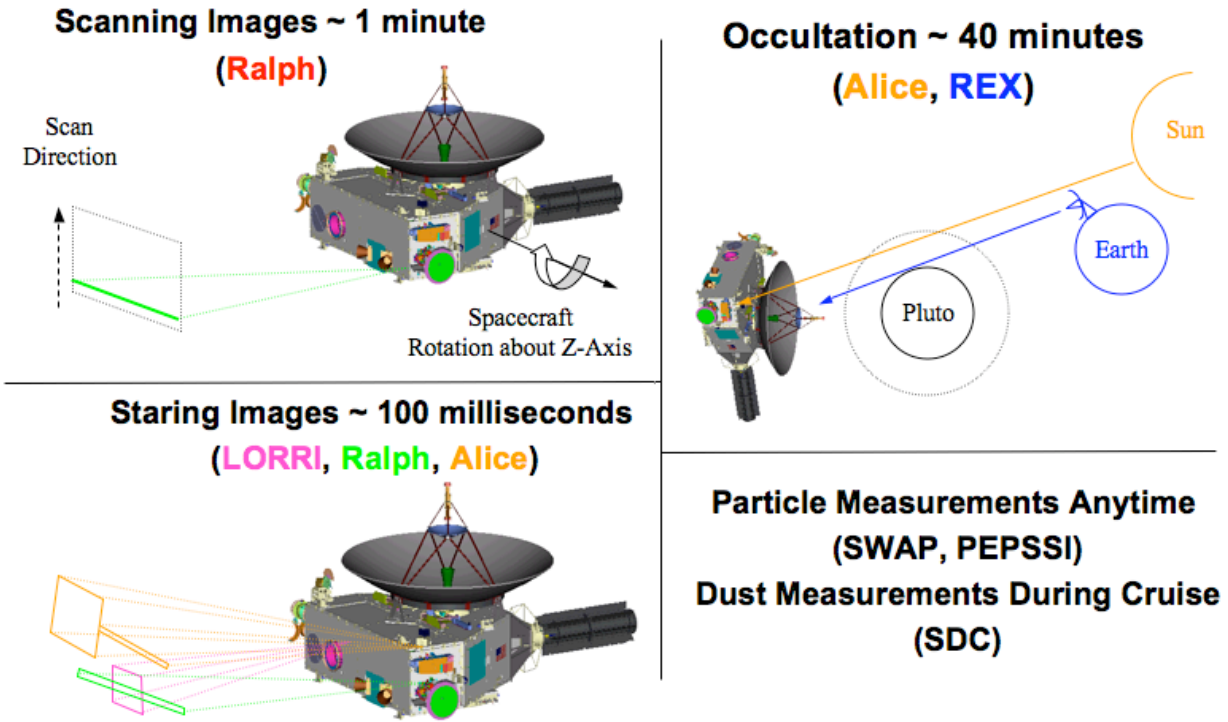

Fig. 4: Types of New Horizons observations. Typical Ralph MVIC Time Delay Integration (TDI) and LEISA observations (upper left) are performed by rotating the spacecraft about the Z-axis. Typical Ralph MVIC frame, LORRI, and Alice airglow observations (lower left) are made with the spacecraft staring in a particular direction. The Alice and REX occultation observations (upper right) are performed by pointing the antenna at the Earth and the Alice occultation channel at the sun, so that radio signals from the DSN on Earth can be received by REX at the same time that Alice observes the Sun. Observations by the particle instruments (SWAP, PEPSSI, and VB-SDC; lower right) can occur essentially anytime, in either spinning or 3-axis mode. However, most of the VB-SDC data will be collected during cruise mode, when the other instruments are in hibernation mode and the spacecraft is passively spinning, because thruster firings add a large background noise level to the VB-SDC's data.

TABLE I

New Horizons Instruments: Pluto System Measurement Objectives and Characteristics (PI=Principal Investigator; Instrument Characteristics are summary values with details provided in the individual instrument papers)

| Instrument, PI | Measurement Objectives | Instrument Characteristics |
|---|---|---|
| UV imaging spectrometer (Alice), S. A. Stern (SwRI), SwRI | • **Upper atmospheric temperature and pressure profiles of Pluto**<br>• **Temperature and vertical temperature gradient should be measured to ~10% at a vertical resolution of ~100 km for atmospheric densities greater than ~$10^9$ cm$^{-3}$.**<br>• **Search for atmospheric haze at a vertical resolution <5 km**<br>• **Mole fractions of $N_2$, CO, $CH_4$ and Ar in Pluto's upper atmosphere.**<br>• **Atmospheric escape rate from Pluto**<br>• Minor atmospheric species at Pluto<br>• Search for an atmosphere of Charon<br>• Constrain escape rate from upper atmospheric structure | UV spectral imaging; 465-1880 Å; FOV 4° x 0.1° plus 2° x 2°; Resolution 1.8 Å/spectral element, 5 mrad/pixel; Airglow and solar occultation channels |
| Multispectral Visible Imaging Camera (Ralph/MVIC), S. A. Stern (SwRI), Ball and SwRI | • **Hemispheric panchromatic maps of Pluto and Charon at best resolution exceeding 0.5 km/pixel**<br>• **Hemispheric 4-color maps of Pluto and Charon at best resolution exceeding 5 km/pixel**<br>• **Search for/map atmospheric hazes at a vertical resolution < 5 km**<br>• High resolution panchromatic maps of the terminator region<br>• Panchromatic, wide phase angle coverage of Pluto, Charon, Nix, and Hydra<br>• Panchromatic stereo images of Pluto and Charon, Nix, and Hydra<br>• Orbital parameters, bulk parameters of Pluto, Charon, Nix, and Hydra<br>• Search for rings<br>• Search for additional satellites | Visible imaging; 400 - 975 nm (panchromatic); 4 color filters (Blue, Red, Methane, Near-IR); FOV 5.7° x 0.15° (stare, pan), or 5.7° x arbitrary (scan); IFOV 20 μrad/pixel |
| Linear Etalon Imaging Spectral Array (Ralph/LEISA), D. Jennings (GSFC), GSFC, Ball, and SwRI | • **Hemispheric near-infrared spectral maps of Pluto and Charon at best resolution exceeding 10 km/pixel**<br>• **Hemispheric distributions of $N_2$, CO, $CH_4$ on Pluto at a best resolution exceeding 10 km/pixel.**<br>• Surface temperature mapping of Pluto and Charon<br>• phase-angle-dependent spectral maps of Pluto and Charon | IR spectral imaging; 1.25 to 2.5 μm; 1.25-2.50 μm, $\lambda/\delta\lambda \approx 240$; 2.10-2.25 μm, $\lambda/\delta\lambda \approx 550$; FOV 0.9° x 0.9°; IFOV 62 μrad/pixel |
| Radio Science Experiment (REX), L. Tyler (Stanford), Stanford and JHU/APL | • **Temperature and pressure profiles of Pluto's atmosphere to the surface**<br>• **Surface number density to ±1.5%, surface temperature to ±2.2 °K and surface pressure to ±0.3 μbar.**<br>• Surface brigtness temperatures on Pluto and Charon (give wavelength)<br>• Masses and chords of Pluto and Charon; detect or constrain J2s.<br>• Detect, or place limits on, an ionosphere for Pluto | X-band (7.182 GHz uplink, 8.438 GHz downlink); Radiometry $T_{Noise}$ < 150 K; Ultra-Stable Oscillator (USO) frequency stability: $\delta f/f = 3 \times 10^{-13}$ over 1 sec |

TABLE I (continued)
New Horizons Instruments: Measurement Objectives and Characteristics

| Instrument, PI, Builder | Measurement Objectives | Instrument Characteristics |
|---|---|---|
| Long Range Reconnaissance Imager (LORRI), A. Cheng (JHU/APL), JHU/APL and SSG | • **Hemispheric panchromatic maps of Pluto and Charon at best resolution exceeding 0.5 km/pixel.**<br>• **Search for atmospheric haze at a vertical resolution <5 km**<br>• Long time base of observations, extending over 10 to 12 Pluto rotations<br>• Panchromatic maps of the far-side hemisphere<br>• High resolution panchromatic maps of the terminator region<br>• Panchromatic, wide phase angle coverage of Pluto, Charon, Nix, and Hydra<br>• Panchromatic stereo images of Pluto, Charon, Nix, and Hydra<br>• Orbital parameters, bulk parameters of Pluto, Charon, Nix, and Hydra<br>• Search for satellites and rings | Visible panchromatic images;<br>350 – 850 nm;<br>FOV 0.29° × 0.29°;<br>IFOV 5 μrad/pixel;<br>Optical Navigation |
| Solar Wind At Pluto (SWAP), D. McComas (SwRI), SwRI | • **Atmospheric escape rate from Pluto**<br>• Solar wind velocity and density, low energy plasma fluxes and angular distributions, and energetic particle fluxes at Pluto-Charon<br>• Solar wind interaction of Pluto and Charon | Solar wind detector<br>FOV 200° x 10°<br>Energy Range 0.25-7.5 keV<br>Energy Resolution<br>  RPA: 0.5 V (< 1.5 keV)<br>  ESA: 0.4 ΔE/E (> 1.4keV) |
| Pluto Energetic Particle Spectrometer Science Investigation (PEPSSI), R. McNutt (JHU/APL), JHU/APL | • Composition and density of pick-up ions from Pluto, which indirectly addresses the atmospheric escape rate<br>• Solar wind velocity and density, low energy plasma fluxes and angular distributions, and energetic particle fluxes in the Pluto system | Energetic particle detector<br>Energy Range 1 kev-1 MeV<br>FOV 160° x 12°<br>Resolution 25° x 12° |
| Venetia Burney Student Dust Counter (VB-SDC), M. Horanyi (U. Colorado), LASP/Colorado | • Trace the density of dust in the Solar System along the New Horizons trajectory from Earth to Pluto and beyond. | 12 PVF panels to detect dust impacts and 2 control panels shielded from impacts |

In the following subsections, we provide further discussion on each of the New Horizons instruments. We attempt to provide a high-level summary of the instruments' capabilities, with detailed descriptions left to the individual instrument papers, which are referenced in each subsection.

## 3.2 ALICE

The Alice instrument aboard New Horizons is an ultraviolet (UV) imaging spectrometer that provides moderate spectral and spatial resolution capabilities over the wavelength range ~465-1880 Å with a peak effective area of ~0.3 cm$^2$. Light enters Alice's f/3 telescope via either the main entrance aperture (called the Airglow Aperture, co-aligned with the Ralph and LORRI apertures), or, via a small, fixed pickoff mirror, through the Solar Occultation Channel (SOCC, co-aligned with the New Horizons high-gain antenna). Light from either aperture is reflected off the 4 cm x 4 cm primary mirror, passes through a single slit, is reflected off a holographic grating, and finally is detected using a photon-counting, microchannel plate double delay line device, read out as a 32 x 1024 element digital array. The SOCC aperture is stopped down by a factor of 6400 relative to the Airglow Aperture to allow Alice to look directly at the Sun for solar occultations of Pluto's and Charon's atmospheres. The Alice entrance slit is a "lollipop" (see Fig. 3) with a 0.1° x 4° "slot" used primarily for airglow observations and a 2° x 2° "box" used mainly during solar occultation observations. The point source spectral resolution is 3-6 Å, depending on wavelength, and the plate scale in the spatial dimension is 0.27° per pixel. During the Pluto and Charon occultation observations, the Sun has an apparent diameter of ~1', and the spectral resolution is 3-3.5 Å. During filled-slit airglow aperture observations, the spectral resolution is ~9-10 Å.

*Alice* is a name, not an acronym, taken from one of the main characters of the American television show *The Honeymooners*. Alice is sometimes called Pluto-Alice (P-Alice) to distinguish it from its predecessor, Rosetta-Alice (R-Alice), which is a similar instrument being flown on the European Space Agency (ESA) Rosetta mission to comet 67P/Churyumov-Gerasimenko. Compared to R-Alice, P-Alice has a somewhat different bandpass and various enhancements to improve reliability. P-Alice also includes a separate solar occultation channel, which is not available on R-Alice. Both P-Alice and R-Alice are significantly improved versions of the Pluto mission "HIPPS" UV spectrograph (HIPPS/UVSC), which was developed at Southwest Research Institute (SwRI) in the mid-1990s with funds from NASA, JPL, and SwRI.

Alice's principal measurement objectives and its key characteristics are summarized in Table I. Alice was designed to measure Pluto's upper atmospheric composition and temperature, which is a New Horizons Group 1 scientific objective. Alice will also obtain model-dependent escape rate measurements from Pluto's atmosphere, and it will provide some limited surface mapping and surface composition capabilities in the UV. Alice's spectral bandpass includes lines of CO, atomic H, Ar, and Ne, which may be detectable as airglow, and the electronic bands of $N_2$, $CH_4$, and other hydrocarbons and nitriles, which are detectable during solar and stellar occultation observations. Young et al. (2007) provide a detailed discussion of Alice's scientific objectives. Stern et al. (2007) should be consulted for further details on Alice's design and performance.

## 3.3 RALPH: MVIC AND LEISA

Ralph and Alice together comprise the primary remote sensing payload on New Horizons. *Ralph* is named after Alice's husband in *The Honeymooners*. It is a combined visible/NIR imager (called MVIC) and imaging IR spectrograph (called LEISA). Both of these two focal planes are fed by a single telescope assembly. *MVIC (Multi-spectral Visible Imaging Camera)* is an optical imager employing CCDs with panchromatic and color filters. *LEISA (Linear Etalon Imaging Spectral Array,) is* a near infrared (IR) imaging spectrograph employing a 256 x 256 mercury cadmium telluride (HgCdTe) array. In addition to its scientific capabilities, MVIC also serves as an Optical Navigation camera for New Horizons.

The common telescope assembly for Ralph has a three-mirror, off-axis anastigmat design with a 7.5 cm primary mirror. A dichroic reflects the optical light to the MVIC focal plane and transmits the IR light to the LEISA focal plane. Only one focal plane is active at a time, with a relay used to select either MVIC or LEISA.

Owing to Ralph's critical role in achieving the New Horizons Group 1 scientific objectives, all of its electronics and some of its focal plane CCDs are redundant.

The MVIC focal plane has seven independent CCD arrays mounted on a single substrate. Figure 3 shows the relative positions of the arrays, as projected on the sky. Six of the arrays have 5000 (columns) x 32 (rows) photosensitive pixels and operate in time-delay integration (TDI) mode. Two of the TDI arrays provide panchromatic (400-975 nm) images, and the other four TDI arrays provide, respectively, color images in blue (400-550 nm), red (540-700 nm), near IR (780–975 nm), and narrow band methane (860–910 nm) channels. The frame transfer array has 5000 (columns) x 128 (rows) pixels and provides panchromatic images (400-975 nm). All of the MVIC arrays have square pixels that are 20 μrad on a side. Thus, the TDI arrays have a field of view of 5.7° x 0.037°, and the frame transfer array has a field of view of 5.7° x 0.15°. To obtain MVIC TDI images, the spacecraft scans the TDI arrays across the target (Fig. 4) at the same rate that charge is shifted from one row to the next, so that the effective exposure time is 32 times the row transfer time. The two TDI panchromatic arrays are sized to meet the 0.5 km/pixel Group 1 mapping requirement near closest approach when Pluto's diameter subtends ~5000 pixels. Each panchromatic array can be operated independently, for redundancy. The four color arrays are operated in tandem. The primary measurement objectives and key characteristics of MVIC are summarized in Table I.

MVIC images in the three broadband colors will provide information on spectral slopes of Pluto's surface and on its atmospheric properties. The narrow band filter permits mapping of the surface methane abundance, as the well-known 890 nm absorption band is the strongest methane feature available at optical wavelengths. The near IR filter doubles as the continuum comparison for this methane mapping. MVIC's framing array is operated in stare, not scanning, mode, and is used when geometrical fidelity is important (e.g., for optical navigation) or when scanning is not practical (e.g., observing Pluto at closest approach when the apparent motion is too fast). The 700-780 nm gap between the red and near IR bandpasses overlaps another methane band at 740 nm; combining data from the panchromatic, blue, red, and near IR filters can provide some information about band depth in this "virtual" filter. Young et al. (2007) discuss MVIC's scientific objectives in more detail. Further details on MVIC and its performance can be found in Reuter et al. (2007).

LEISA's dispersive capability is provided by its wedged etalon (a linear variable filter, or LVF), which is mounted ~100 μm above its 256x256 pixel HgCdTe PICNIC array. The etalon covers 1.25-2.5 μm, a spectral region populated with many absorption features of $N_2$, $CH_4$, $H_2O$, $NH_3$, $CO$, and other molecules, at a resolving power of ~250. A higher-resolution sub-segment, covering 2.10-2.25 μm at a resolving power of ~560, will be used to discern grain sizes, mixing states, and pure versus solid-solution abundances (Quirico et al. 1999). The higher-resolution segment is also critical for taking advantage of the temperature-sensitive $N_2$ bands (Grundy et al. 1993, 1999), and the symmetric, doubled $v_2 + v_3$ $CH_4$ band that is diagnostic of pure versus diluted $CH_4$ abundances (Quirico & Schmitt 1997). As was the case for MVIC, LEISA images are obtained by scanning its field of view across the target (Fig. 4) with the frame transfer rate synchronized with the scan rate. The LVF is oriented so that wavelength varies along the scan direction. Thus, scanning LEISA over a target produces images at different wavelengths (this is unlike the case for MVIC where the scanning simply increases the signal 32-fold). LEISA builds up a conventional spatial-spectral data cube (256 monochromatic images) by scanning the FOV across all portions of the target at a nominal scan rate of 125 μrad/sec. A nominal framing rate of 2 Hz is currently planned to maintain <1 pixel attitude smear and provide good signal-to-noise ratio in the Pluto system. The primary measurement objectives and key characteristics of LEISA are summarized in Table I. Reuter et al. (2007) provide further details on LEISA's design and performance, and Young et al. (2007) provides a more in-depth discussion of LEISA's scientific objectives.

## 3.4 REX

*REX* is the radio science package on New Horizons. *REX* stands for Radio EXperiment. The REX instrument is unique among the suite of instruments comprising the New Horizons payload in that it is physically and functionally incorporated within the spacecraft telecommunications subsystem. Because this system is entirely redundant, two REX's are carried on New Horizons. They can be used simultaneously to increase SNR.

The REX principle of operations for radio science is as follows: the 2.1 m High Gain Antenna aboard New Horizons (see Fig. 1) receives radio signals from NASA's Deep Space Network (DSN) at a carrier frequency of 7.182 GHz. New Horizons transmits radio signals via the antenna to the DSN at a carrier frequency of 8.438 GHz. By measuring phase delays in the received signal as a function of time, the instrument allows one to invert a radio occultation profile into a temperature, number density profile of the intervening atmosphere, if it is sufficiently dense. REX can also operate in a passive radiometry mode to measure radio brightness temperatures at its receiver frequency.

The heart of the REX instrument is an ultra-stable oscillator (USO), which operates at 30 MHz for the down-conversion to an intermediate frequency (IF). An Actel Field Programmable Gate Array (FPGA) takes samples of the IF receiver output and generates wideband radiometer and narrowband sampled signal data products. The REX hardware also includes an analog-to-digital converter (ADC) and other electronics interface components in the telecommunications system. As noted above, there are two copies of the entire telecommunications system (except for the High Gain Antenna), which means that there is full system redundancy in the REX capabilities. The primary measurement objectives and key characteristics of REX are summarized in Table I. Tyler et al. (2007) discuss REX and its performance in much greater detail.

REX addresses the Group 1 scientific objective of obtaining Pluto's atmospheric temperature and pressure profiles down to the surface using a unique uplink radio occultation technique. REX detects the changes induced by Pluto's atmosphere in the radio signal transmitted to the spacecraft from the DSN. This differs from the typically used downlink method, in which the spacecraft transmits to receivers on Earth. REX will also address Group 2 and Group 3 scientific objectives, including probing Pluto's ionospheric density, searching for Charon's atmosphere, refining bulk parameters like mass and radius, and measuring the surface emission brightness at a wavelength of 4.2 cm, which permits the determination of both the dayside and nightside brightness temperatures with an angular resolution of ~1.2° (full-width between the 3 dB points). Young et al. (2007) provide further discussion of the REX's scientific objectives.

## 3.5 LORRI

The *Long Range Reconnaissance Imager (LORRI)* is a narrow angle (field of view=0.29° x 0.29°), high resolution (4.96 μrad/pixel), panchromatic (350-850 nm) imaging system. It was placed on New Horizons to augment and also back up Ralph's panchromatic imaging capabilities. LORRI's primary function is to provide higher resolution imagery.

LORRI's input aperture is 20.8 cm in diameter, making LORRI one of the largest telescopes flown on an interplanetary spacecraft. The large aperture, in combination with a high throughput ($QE_{peak} \approx 60\%$) and wide bandpass, will allow LORRI to achieve a signal-to-noise ratio exceeding 100 during disk-resolved observations of Pluto, even though exposure times must be kept below 100 ms to prevent smearing from pointing drift. A frame transfer 1024 x 1024 pixel (optically active region), thinned, backside-illuminated charge-coupled device (CCD) detector records the image in the telescope focal plane. The CCD output is digitized to 12 bits and stored on the spacecraft's solid state recorder (SSR).

Raw images can be downlinked, but typically the images will be either losslessly or lossy compressed before transmission to the ground in order to minimize the use of DSN resources. LORRI image exposure times can be varied from 0 ms to 29,967 ms in 1 ms steps, and images can be accumulated at a maximum rate of 1 image per second. LORRI's large dynamic range allows it to be an imaging workhorse during the Jupiter encounter, when saturation limits MVIC observations to relatively large solar phase angles.

LORRI operates in an extreme thermal environment, mounted inside the warm spacecraft and viewing cold space, but the telescope's monolithic, silicon carbide construction allows the focus to be maintained over a large temperature range (-120 C to 50 C) without any focus adjustment mechanisms. Indeed, LORRI has no moving parts making it a relatively simple, reliable instrument that is easy to operate. A one-time deploy aperture door, mounted on the spacecraft structure, protected LORRI from the harsh launch environment. Cheng et al. (2007) provide a detailed description of LORRI and its performance.

Owing to its higher spatial resolution, higher sensitivity, and lower geometrical distortion (< 0.5 pixel across the entire field of view) compared to Ralph/MVIC, LORRI is also serving as the prime optical navigation instrument on New Horizons. During a typical 100 ms exposure using the full format (1024 x 1024) mode, LORRI can achieve a signal-to-noise ratio of ~5 on V=13 stars. On-chip 4x4 binning, used in conjunction with a special pointing control mode that permits exposing up to 10 s while keeping the target within a single rebinned pixel, allows imaging of point sources as faint as V≈18, which will permit LORRI to detect a 50 km diameter KBO ~7 weeks prior to encounter, thereby enabling accurate targeting to the KBO.

LORRI's primary measurement objectives and key characteristics are summarized in Table I. LORRI has first successfully detected Pluto (on September 21, 2006 at a distance of 28 AU), and its resolution at Pluto will start exceeding that available from the Hubble Space Telescope approximately 3 months prior to closest approach. En route to Pluto, LORRI will obtain rotationally resolved phase curves of Pluto and later Charon, once the two can be separately resolved. LORRI will obtain panchromatic maps over at least 10 Pluto rotations during approach, with the final complete map of the sunlit hemisphere exceeding a resolution of 0.5 km/pixel. LORRI will map small regions near Pluto's terminator with a resolution of ~50 m/pixel near the time of closest approach, depending on the closest approach distance selected. LORRI will also be heavily used for studies requiring high geometrical fidelity, such as the determining the shapes of Pluto, Charon, Nix, and Hydra and refining the orbits of all these objects relative to the system barycenter. LORRI observations at high phase angles will provide a sensitive search for any particulate hazes in Pluto's atmosphere. Young et al. (2007) provides a more detailed discussion of the scientific objectives addressed by LORRI observations.

## 3.6 SWAP

The Solar Wind Around Pluto (SWAP) instrument is one of two particle detection in situ instruments aboard New Horizons. It is comprised of a retarding potential analyzer (RPA), a deflector (DFL), and an electrostatic analyzer (ESA). Collectively, these elements are used to select the angles and energies of solar wind ions entering the instrument. The selected ions are directed through a thin foil into a coincidence detection system: the ions themselves are detected by one channel electron multiplier (CEM), and secondary electrons produced from the foil are detected by another CEM. SWAP can measure solar wind particles in the energy range from 25 eV up to 7.5 keV with a resolution of $\Delta E/E <0.4$. SWAP has a fan-shaped field of view that extends >200º in the XY-plane of the spacecraft by >10º out of that plane (see Fig. 2). For typical observations, SWAP measures solar wind speed and density over a 64 sec measurement cycle. The principal measurement objectives and key characteristics of SWAP are summarized in Table I. Further details on SWAP and its performance can be found in McComas et al. (2007).

SWAP was designed to measure the interaction of the solar wind with Pluto, which addresses the Group 1 scientific objective of measuring Pluto's atmospheric escape rate. Additionally, SWAP has a specific goal of characterizing the solar wind interaction with Pluto as a Group 2 objective. SWAP also addresses the Group 3 objectives of characterizing the energetic particle environment of Pluto and searching for magnetic fields, which it does indirectly. For more details on SWAP's scientific objectives, see McComas et al. (2007) and Young et al. (2007).

## 3.7 PEPSSI

The *Pluto Energetic Particle Spectrometer Science Investigation (PEPSSI)* is the other in situ particle measurement instrument aboard New Horizons. It is a compact, radiation-hardened particle instrument comprised of a time-of-flight (TOF) section feeding a solid-state silicon detector (SSD) array. Each SSD has 4 pixels, 2 dedicated to ions, and 2 for electrons. PEPSSI's field of view (FOV) is fan-like and measures 160º x 12º, divided into six angular sectors of 25º x 12º each. Ions entering the PEPSSI FOV generate secondary electrons as they pass through entrance and exit foils in the TOF section, providing "start" and "stop" signals detected by a microchannel plate (MCP). Particle energy information, measured by the SSD, is combined with TOF information to identify the particle's composition. Each particle's direction is determined by the particular 25° sector in which it is detected. Event classification electronics determine incident mass and energy, with 12 channels of energy resolution. Protons can be detected in the energy range 40-1000 keV, electrons in the range 25-500 keV, and CNO ions in the range 150-1000 keV. TOF-only measurements extend to <1 keV for protons, to 15 keV for CNO ions, and to 30 keV for $N_2^+$. TOF measurements are possible in the range 1-250 ns to an accuracy of ±1 ns. The geometrical factor for ions is slightly larger than 0.1 cm$^2$ sr. A typical measurement includes 8-point spectra for protons and electrons and reduced resolution energy spectra for heavier ions for all six look directions. The mass resolution of PEPSSI varies with energy: for CNO ions, it is <5 AMU for >1.7 keV/nucleon, and <2 AMU for >5 keV/nucleon. The principal measurement objectives and key instrument characteristics of PEPSSI are summarized in Table I. McNutt et al. (2007) provide a detailed discussion of PEPSSI and its performance.

The PEPSSI design is derived from that of the Energetic Particle Spectrometer (EPS), which is flying on the MESSENGER mission to Mercury. PEPSSI has thinner foils than EPS, which enables measurements down to smaller energy ranges. PEPSSI also has a slightly increased geometric factor and draws less power than EPS. Both EPS and PEPSSI trace back their heritage to a NASA PIDDP program in the 1990s to develop a particle instrument for use on a Pluto flyby mission.

By measuring energetic pickup ions from Pluto's atmosphere, PEPSSI provides information related to the atmospheric escape rate on Pluto, which is a New Horizons Group 1 scientific objective. PEPSSI's primary role, however, is to address the Group 3 objective of characterizing the energetic particle environment of Pluto. Fluxes of energetic pickup ions may be measured as far as several million kilometers from Pluto (see Bagenal et al. 1997), and PEPSSI observations will be used to determine the mass, energy spectra, and directional distributions of these energetic particles (Bagenal & McNutt 1989). Secondarily, PEPSSI will also provide low-resolution, supporting measurements of the solar wind flux, complementing SWAP. Young et al. (2007) provides a more detailed discussion of PEPSSI's scientific objectives.

## 3.8 VB-SDC

The Student Dust Counter (SDC), also known as the Venetia Burney SDC in honor of the student who named Pluto in 1930, is an impact dust detector that will be used to map the spatial and size distribution of interplanetary dust along the trajectory of the New Horizons spacecraft from the inner solar system to and through the Kuiper Belt..

Unlike all of the other instruments, the VB-SDC was not part of the original New Horizons proposal and was added by NASA as an Education and Public Outreach (EPO) experiment. For the first time ever, students were given the opportunity to design, build, and operate an instrument for an interplanetary mission. (NASA-certified personnel performed all quality assurance inspections and supervised the final assembly.) Approximately 20 undergraduate physics and engineering students at the University of Colorado worked on the VB-SDC and, despite getting a rather late start, their instrument was the first to be delivered to the New Horizons spacecraft.

The VB-SDC's sensors are thin, permanently polarized polyvinylidene fluoride (PVDF) plastic films that generate an electrical signal when dust particles penetrate their surface. The SDC has a total sensitive surface area of ~0.1 m$^2$, comprised of 12 separate film patches, each 14.2 cm x 6.5 cm, mounted onto the top surface of a support panel. In addition, there are two reference sensor patches mounted on the backside of the detector support panel, protected from any dust impacts. These reference sensors, identical to the top surface sensors, are used to monitor the various background noise levels, from mechanical vibrations or cosmic ray hits.

The entire support panel is mounted on the exterior of the New Horizons spacecraft, outside the spacecraft multi-layer insulating (MLI) blanket, facing the ram (-Y) direction. The VB-SDC observations are most useful during the cruise phases of the mission, when the spacecraft is spinning and the other instruments are turned off. Thruster firings during 3-axis operations generate large VB-SDC background signals, which make it very difficult to detect true IDP impacts.

The VB-SDC was designed to resolve, to within a factor of ~2, the masses of interplanetary dust particles (IDPs) in the range of $10^{-12} < m < 10^{-9}$ g, which corresponds roughly to a size range of 1 – 10 μm in particle radius. Bigger grains are also recorded, but their masses cannot be resolved. With the characteristic spacecraft speed during cruise of ~13 km/s, current models of the dust density in the solar system (Divine, 1993) suggest that the VB-SDC should record approximately 1 IDP hit per week.

The principal measurement objectives and key instrument characteristics of the VB-SDC are summarized in Table I. Horanyi et al. (2007) provide a detailed discussion of the VB-SDC and its performance.

# 4. Science Payload Commissioning Overview

The New Horizons instrument commissioning activities began shortly after the nominal performance of the spacecraft subsystems was verified; this was approximately 1 month after launch. Over a period of about 8 months, each instrument team developed a detailed set of science activity plans (SAPs) to characterize the in-flight performance and functionality, and to verify that their measurement objectives could be achieved. Functional tests were executed first to demonstrate that critical engineering parameters (e.g., currents, voltages, temperatures, etc.) fell within their expected ranges. After nominal functional performance was verified, a series of performance tests were executed for each instrument. All of the commissioning tests discussed below took place during calendar year 2006. A small subset of commissioning activities (about 10% of the total) remain to be completed during and after the Jupiter encounter in early 2007.

The instruments completed their functional tests during February-March 2006. The first observations of an external target are termed "first light" observations, and these were staggered throughout the May to September 2006 period for the various instruments. Alice detected interplanetary hydrogen Lyman-$\alpha$ and Lyman-$\beta$ emission during its first light observations on May 29. Alice then observed two UV calibration stars, $\gamma$ Gruis and $\rho$ Leonis, on August 31. Owing to safety reasons, the Alice SOCC door will not be opened until at least March 2007, after the Jupiter encounter. Ralph/MVIC first light occurred during observations of its stellar calibration targets (the M6 and M7 galactic open clusters) through its windowed door on May 10, and then through the opened door on May 28. Both Ralph/MVIC and Ralph/LEISA observed the asteroid 2002 JF56 in a moving target tracking test during May 11-13. Ralph/LEISA made the first observations of its calibration star (Procyon) on June 29. LORRI first light occurred when it opened its aperture door on August 29 and observed M7. LORRI observed M7 for an extensive set of calibration observations on September 3, including a simultaneous observation with MVIC to measure the relative alignments of those two instruments. Both Ralph and LORRI observed Jupiter on September 4 as test observations in preparation for the Jupiter encounter in February 2007. Ralph and LORRI also observed Uranus and Neptune in September for optical navigation testing. LORRI observed Pluto during observations on September 21 and 23, and the Jovian irregular satellite Himalia on September 22, again as part of optical navigation testing. The first use of REX mode by the telecommunications system took place on April 19. REX scanning observations to measure the high gain antenna (HGA) beam pattern were performed on June 20. Two radio calibration sources (Cass A and Taurus A) and "cold sky" were observed on June 29 to measure the REX radiometry mode performance. All REX calibration observations in 2006 were performed on side-A; side-B calibration observations were executed in early January 2007. SWAP's door was opened on March 13, but the first solar wind observations started in late-September and continued through December. PEPSSI's door was opened on May 3, but its ability to measure particles was first tested in June. The VB-SDC attempted to take science data in early-March, but the spacecraft was in 3-axis mode and the high VB-SDC background rate produced by the nearly continual thruster firings made it essentially impossible to detect real dust particle events. The VB-SDC had its first real chance to detect dust particles while the spacecraft was in "passive" spin mode (thruster firings still occur during "active" spin mode) in April, but the relatively low count rate expected requires that the instrument be well-calibrated and the data carefully analyzed, Additional VB-SDC data was then taken from October through December 2006, while the spacecraft remained in spin mode.

# 5. In Flight Hibernation, Annual Checkouts, and Encounter Rehearsals

The New Horizons mission is exceptionally long in duration, with the primary mission objective not being completed until nearly 10 years after launch. Activities during the mission are generally either front-end or back-end loaded, with the first 14 months busy with instrument commissioning and the Jupiter encounter, and the last year of the mission devoted to intensive observations of the Pluto system. For most of the time during the 8 years between the encounter phases (2007-2014, inclusive), the spacecraft will be placed into a "hibernation" mode, with all non-essential subsystems, including the scientific payload, powered off. This preserves component life.

During the hibernation period, beacon radio tones are sent periodically from the spacecraft to the Earth that allow flight controllers to verify the basic health and safety of the spacecraft. Additionally, monthly telemetry passes are scheduled to collect engineering trend data.

Although the spacecraft is kept in hibernation to reduce component use prior to the Pluto-system encounter, it is also important to verify periodically the performance of the spacecraft subsystems and instruments, and to keep the mission operations team well trained and prepared for the Pluto encounter activities. Therefore, the spacecraft will be brought out of hibernation each year for roughly 60 days, called "annual checkouts" (ACOs), during which time the performance of the spacecraft subsystems and instruments can be verified. Generally, the ACO instrument activities are comprised of a subset of the commissioning activities that focus on the instrument's performance (e.g., stellar calibration observations). ACOs are also the opportunity for annual cruise science observations to be collected, such as interplanetary charged particle measurements, studies of the hydrogen distribution in the interplanetary medium, and extensive phase curve studies of Pluto, Charon, Uranus, Neptune, Centaurs, and KBOs, none of which can be obtained from spacecraft near Earth.

In addition, two full rehearsals of the Pluto encounter will be conducted, during the summers of 2012 and 2014, respectively, that will serve both to verify that the Pluto encounter sequence will work and to provide essential training for the mission operations team in preparation for the actual encounter.

## 6. Current Status of the Science Payload

All seven of the instruments comprising the New Horizons science payload have essentially completed their in-flight commissioning activities, with only a few tests remaining to be executed. In all cases, the in-flight performance verifies that the science payload can meet its measurement objectives, thereby accomplishing all of the scientific objectives of the New Horizons mission.

The Ralph instrument was used to observe asteroid 2002 JF56 during a serendipitous flyby at a closest approach distance of 100,000 km on 2006 June 13 (Olkin et al. 2006), which verified the spacecraft's ability to track reliably a fast-moving target. The SWAP, PEPSSI, and VB-SDC instruments also began taking scientific data, in addition to commissioning data, during 2006.

All of the instruments will participate in the upcoming encounter with Jupiter (with closest approach on 2007 February 28), which will be considerably more ambitious than any of the activities executed to date. In fact, the Jupiter encounter will likely have approximately double the number of observations, and double the data volume, compared to what is currently planned for the Pluto encounter in 2015. Most importantly, however, the Jupiter encounter provides an invaluable and unique opportunity to test the mission's capabilities. Even if some of the observations taken in the Jovian system fail, the lessons learned from that encounter will undoubtedly improve the prospects for a successful Pluto system encounter in 2015, which is the most important activity of the New Horizons mission.

## Acknowledgments

We thank all of the New Horizons Instrument Teams for their extraordinary efforts in designing, developing, testing, and delivering a highly capable science payload that promises to revolutionize our understanding of the Pluto system and the Kuiper belt. We also thank the numerous contractors who partnered with the Instrument Teams for their outstanding work and dedication. Partial financial support for this work was provided by NASA contract NAS5-97271 to the Johns Hopkins University Applied Physics Laboratory.